\documentclass[12pt]{article}%
\usepackage{amsmath}
\usepackage{graphicx}%
\usepackage{amsfonts}%
\usepackage{amssymb}
\begin{document}

\title{Areal Theory\thanks{Based on a talk given at the ``Coral Gables'' conference,
Fort Lauderdale, Florida, 12 December 2001. \copyright\ American Institute of
Physics, http://proceedings.aip.org/proceedings/ \#624: Coral Gables
Conference on Cosmology and Elementary Particle Physics; B. N. Kursunoglu, Editor.}}
\author{{\large Thomas Curtright}\\{\normalsize University of Miami, Coral Gables, Florida 33124}\\{\small curtright@physics.miami.edu}}
\date{}
\maketitle

\begin{abstract}
New features are described for models with multi-particle area-dependent
potentials, in any number of dimensions. \ The corresponding many-body field
theories are investigated for classical configurations. \ Some explicit
solutions are given, and some conjectures are made about chaos in such field theories.

\bigskip

\bigskip

\end{abstract}

Area-dependent potentials, $V\left(  A=\mathbf{r}_{1}\wedge\mathbf{r}%
_{2}\right)  $, or their mathematical equivalents, appear in several physical
problems of contemporary interest. \ Notable among these problems are
Yang-Mills theory (especially for spatially homogeneous configurations, with
only time dependence) \cite{Chang}, extended supersymmetric field theories
with (pseudo)scalar self-interactions \cite{Howe,Ferrara,Lee}, and more
recently, membrane models \cite{de Wit,Graf}. \ In the last century, Feynman
\cite{Feynman} even assigned to students an exercise involving such
potentials: \ Show the energy spectrum for a quantized areal potential model
is discrete and quantum particles cannot escape from such a potential, even
though the spectrum for the classical model is continuous\ and classical
particles can escape along special trajectories for which $A=0$. \ Many of us
pondered this problem over the intervening years, especially in the field
theory context \cite{TLC1989}. \ Meanwhile, Barry Simon worked out five or six
solutions to Feynman's exercise and published them \cite{Simon}.

Classically, such potential models are interesting insofar as they may provide
simple examples of chaotic systems \cite{Saviddy,Carnegie}. \ While there are
special trajectories for which the motion is quite regular, including those
for which particles can escape (such as straight line, free particle motion),
for most trajectories this is not the case. \ The current consensus is such
models are not integrable in the Liouville sense, and do not admit the
construction of a Lax pair. \ Is this really so?

Quantum mechanically, there are obviously more interesting questions to ask
about such potential models than simply whether the energy spectrum is
discrete. \ In particular, just what \emph{is}\ that discrete spectrum? \ Is
the quantized model, or some simple variant of it, completely integrable
\cite{Jaffe}, even if the classical is not? \ Do quantum effects sufficiently
ameliorate any chaotic classical trajectories \cite{Gutzwiller} to permit
closed-form solutions for the wave functions, or exotic but useful forms
\cite{Cataplex} for the propagator?

With these questions in mind, I will describe in this talk\footnote{In the
actual talk, I began with a simple model of $N$ distinguishable point
particles on the plane, with $V=A$. \ This is based on work in collaboration
with Alexios Polychronakos and Cosmas Zachos \cite{CPZ}. \ We have completely
understood this model's classical and quantum properties. \ Since the
potential is quadratic, this is not very difficult to do, in principle, but
nevertheless it was necessary to develop an efficient formalism to handle an
arbitrary number of particles. \ We have done this using so-called cyclotomic
coordinates that have previously been used to analyze toroidal membrane
dynamics \cite{FZ}. \ I view the analysis of this $V=A$ model as a precursor
to point particle models with $V=A^{2}$. \ These latter models are much more
interesting, but are not so easy to solve. \ While I have made a modicum of
progress in an unfinished attempt to construct a Lax pair for certain variants
of $V=A^{2}$ models, in the point particle approach \cite{Perelomov}, I will
not discuss this here. \ Instead, I will shift from point particle models to
field theories.} some new features of models with area-dependent potentials.
\ As non-relativistic many-body field theories, I believe that area-dependent
models are very interesting. \ I will argue this beginning with the classical
field versions. \ I explicitly solve these for some special configurations,
and then investigate general situations. \ The classical field theory problem
always reduces to solving the linear Schr\"{o}dinger equation in a
self-consistent effective potential. \ In general, the effective potential for
the $V=A$ model is anisotropic and linear, while for the $V=A^{2}$ model it is
anisotropic and quadratic. \ In both cases the effective potential is
determined from the initial data by a closed set of coupled time-dependent
equations. \ I will analyze these equations, especially for the case of
spherically symmetric particle density, and I will solve them in that special
situation. \ However, I have not completely solved the area-dependent models
for general, anisotropic initial data. \ Much more work is needed for the
$V=A^{2}$ model.\ \ Additional work is also needed to construct the quantum
versions of these field theories. \ These particular models seem to be ripe
for exploration using deformation quantization \cite{Bayen,CFZ,CUZ} and ideas
from non-commutative geometry \cite{Gracia}.

To begin, we construct a many-body field theory on the plane with a three-body
potential that is just the \emph{signed} area of the three-body triangle
(\emph{not} the absolute value of the area). \ We take a collection of
\emph{three types of non-identical particles}, i.e. three particle
``species'', each represented by its own local field $\psi_{a}$ , $a=1,2,3$.
\ In this case, the Hamiltonian has no lower bound for configurations with
large but negative $A$, so the model is reminiscent of a linear potential
single-particle QM. \ (Actually there is more than reminiscing going on here
as we shall soon see.) \ Explicitly, we define a multi-particle configuration
by the field theory Lagrangian%
\[
L=\sum_{a=1,2,3}i\int\left(  d\mathbf{r}\right)  \psi_{a}^{\ast}\left(
\mathbf{r}\right)  \frac{\partial}{\partial t}\psi_{a}\left(  \mathbf{r}%
\right)  -H\;.
\]
It should be understood that all the fields also depend on a common time, $t$,
but usually we will not explicitly indicate this. \ We then use the obvious
symmetrical Hamiltonian%
\begin{align*}
H  &  =%
{\textstyle\int}
\left(  d\mathbf{r}\right)  \sum_{a=1,2,3}\psi_{a}^{\ast}\left(
\mathbf{r}\right)  \left[  \tfrac{-\nabla^{2}}{2m}\right]  \psi_{a}\left(
\mathbf{r}\right) \\
&  +k%
{\textstyle\iiint}
\left(  d\mathbf{r}_{1}\right)  \left(  d\mathbf{r}_{2}\right)  \left(
d\mathbf{r}_{3}\right)  \;A\left(  \mathbf{r}_{1},\mathbf{r}_{2}%
,\mathbf{r}_{3}\right)  \;\left|  \psi_{1}\left(  \mathbf{r}_{1}\right)
\psi_{2}\left(  \mathbf{r}_{2}\right)  \psi_{3}\left(  \mathbf{r}_{3}\right)
\right|  ^{2}%
\end{align*}
involving $A$, (twice) the area of the triangle formed by the three particles.%
\[
A\left(  \mathbf{r}_{1},\mathbf{r}_{2},\mathbf{r}_{3}\right)  =\mathbf{r}%
_{1}\wedge\mathbf{r}_{2}+\mathbf{r}_{2}\wedge\mathbf{r}_{3}+\mathbf{r}%
_{3}\wedge\mathbf{r}_{1}\;.
\]
This may be rewritten (exactly) as%
\[
H=\sum_{a=1,2,3}%
{\textstyle\int}
\left(  d\mathbf{r}\right)  \psi_{a}^{\ast}\left(  \mathbf{r}\right)  \left[
\tfrac{-\nabla^{2}}{2m}\right]  \psi_{a}\left(  \mathbf{r}\right)  +\tfrac
{1}{3}\sum_{a=1,2,3}%
{\textstyle\int}
\left(  d\mathbf{r}\right)  \left(  S_{a}\left[  \psi\right]  +\mathbf{V}%
_{a}\left[  \psi\right]  \cdot\mathbf{r}\right)  \left|  \psi_{a}\left(
\mathbf{r}\right)  \right|  ^{2}%
\]
where we have introduced the combinations $S_{a}\left[  \psi\right]
+\mathbf{V}_{a}\left[  \psi\right]  \cdot\mathbf{r}$ for $a=1,2,3$ to serve as
\emph{effective potentials} for the three fields. \ These effective potentials
are defined in terms of the fields as%

\begin{align*}
V_{a}^{i}\left[  \psi\right]   &  =\tfrac{1}{2}k\varepsilon^{ij}%
\sum_{b,c=1,2,3}\varepsilon^{abc}\iint\left(  d\mathbf{r}_{2}\right)  \left(
d\mathbf{r}_{3}\right)  \;\left(  \mathbf{r}_{2}-\mathbf{r}_{3}\right)
^{j}\;\left|  \psi_{b}\left(  \mathbf{r}_{2}\right)  \psi_{c}\left(
\mathbf{r}_{3}\right)  \right|  ^{2}\\
S_{a}\left[  \psi\right]   &  =\tfrac{1}{2}k\varepsilon^{abc}\iint\left(
d\mathbf{r}_{2}\right)  \left(  d\mathbf{r}_{3}\right)  \;\left(
\mathbf{r}_{2}\wedge\mathbf{r}_{3}\right)  \;\left|  \psi_{b}\left(
\mathbf{r}_{2}\right)  \psi_{c}\left(  \mathbf{r}_{3}\right)  \right|  ^{2}%
\end{align*}
Note that $S_{a}$ and $\mathbf{V}_{a}$ do \emph{not} depend on $\psi_{a}$ but
\emph{do} depend on the other two field configurations, $\psi_{b\neq a}$.
\ Further observe that $S_{a}\left[  \psi\right]  =0$ if \emph{either} of the
two $\psi$'s involved are of definite parity (either even or odd functions of
$\mathbf{r}$), and $V_{a}^{i}\left[  \psi\right]  =0$ if \emph{both} of the
two $\psi$'s involved are of definite parity. \ Thus the special
configurations where all three species have definite parity reduce to free
fields! \ Again, this is a model on the plane, so each species coordinate
$\mathbf{r}_{a}$ is a two-vector. \ To construct a potential linear in the
signed area in higher dimensions, we would need to take an inner product of
$A\left(  \mathbf{r}_{1},\mathbf{r}_{2},\mathbf{r}_{3}\right)  $ with some
constant $2$-form.

Like the point-particle theory in \cite{CPZ}, this classical field theory is
always \emph{solvable}, even in the general situation when all the fields do
\emph{not} have definite parity. \ Unlike \cite{CPZ}, we work in real
coordinates here. \ The field equations in terms of the effective potentials
are then:%
\[
i\frac{\partial}{\partial t}\psi_{a}\left(  \mathbf{r},t\right)  +\frac{1}%
{2m}\nabla^{2}\psi_{a}\left(  \mathbf{r},t\right)  =\left(  S_{a}\left[
\psi\right]  +\mathbf{V}_{a}[\psi]\cdot\mathbf{r}\right)  \psi_{a}\left(
\mathbf{r},t\right)  \text{ \ \ \ \ no sum }a\text{.}%
\]
For each species we have just Schr\"{o}dinger's equation with a linear (in
$r^{i}$) potential, albeit a non-isotropic linear potential, in general, with
non-constant (in $t$) configuration-dependent coefficients. \ From the above
definitions we obtain
\[
S_{a}\left[  \psi\right]  =\tfrac{1}{2}k\sum_{\substack{b,c=1,2,3\\i,j=1,2}%
}\varepsilon^{ij}\varepsilon^{abc}N_{b}R_{b}^{i}N_{c}R_{c}^{j}\;,\;\;\;V_{a}%
^{i}\left[  \psi\right]  =k\sum_{\substack{b,c=1,2,3\\j=1,2}}\varepsilon
^{ij}\varepsilon^{abc}N_{b}R_{b}^{j}N_{c}%
\]%
\begin{align*}
N_{a}R_{a}^{j} &  \equiv\int\left(  d\mathbf{r}\right)  \,r^{j}\,\left|
\psi_{a}\left(  \mathbf{r},t\right)  \right|  ^{2}\;,\;\;\;N_{a}P_{a}%
^{j}\equiv-i\int\left(  d\mathbf{r}\right)  \,\psi_{a}^{\ast}\left(
\mathbf{r},t\right)  \overleftrightarrow{\nabla}^{j}\psi_{a}\left(
\mathbf{r},t\right)  \;,\\
N_{a} &  \equiv\int\left(  d\mathbf{r}\right)  \,\left|  \psi_{a}\left(
\mathbf{r},t\right)  \right|  ^{2},
\end{align*}
with no sum over $a$ in any of these. \ While these are indeed
configuration-dependent coefficients, the implicit field dependence of the
coefficients is completely tractable. \ From the field equations, the
coefficients in the effective potential obey simple first order
time-derivative equations:%
\begin{align*}
\frac{d}{dt}N_{a} &  =0\;,\;\;\;\frac{d}{dt}R_{a}^{k}=\frac{1}{2m}\,P_{a}%
^{k}\;,\;\\
\frac{d}{dt}P_{a}^{j} &  =-2V_{a}^{j}=-2k\varepsilon^{ij}\sum_{b,c=1,2,3}%
\varepsilon^{abc}N_{b}N_{c}R_{b}^{j}\;,\;\;\;\frac{d}{dt}\left(
\sum_{a=1,2,3}N_{a}P_{a}^{j}\right)  =0\;.
\end{align*}
The last equation represents conservation of the system's total momentum,
while the next to last equation shows that the individual particle species
momenta are not separately conserved, in general.

These first order equations combine to yield linear second order equations
with \emph{constant} coefficients for any given initial data (hence easily
solved).%
\[
m\frac{d^{2}}{dt^{2}}R_{a}^{i}=k\,\sum_{j,b}M_{\left(  ia\right)  \left(
jb\right)  }R_{b}^{j}\;,\;\;\;\;\;M_{\left(  ia\right)  \left(  jb\right)
}=\varepsilon^{ij}\sum_{c=1,2,3}\varepsilon^{acb}N_{c}N_{b}%
\;\;\;\;\;\text{\ (no sum }b\text{)\ .}%
\]
The time-\emph{in}dependent eigenvalues of $M$ are $0$ and $\pm\sqrt
{N_{1}N_{2}N_{3}}\sqrt{N_{1}+N_{2}+N_{3}}$, with each of these three
possibilities occurring twice. \ (Note how the three point-particle mechanics
eigenvalues in \cite{CPZ}\ are obtained by setting all the $N$'s equal to
one.) \ Therefore the $R$'s, and hence all the terms in the effective
potentials, can be solved for in terms of the initial data. \ The three
independent classes of solutions for the $R$'s, corresponding to the three
eigenvalues of $M,$ are functions linear in $t$, real exponentials, and
oscillations. \ The time dependencies of the terms in the effective potential
then follow from their expression in terms of the $R$'s. \ Finally, to
complete the solution, the linear Schr\"{o}dinger equations for the individual
fields must be solved using the solutions for the effective potentials.

I believe, and assert without proof, that this last step can always be carried
out, although there is some possibility of unusual behavior for the fields
because the potential terms in the Schr\"{o}dinger equation are combinations
of polynomials and complex exponentials in $t$. \ Technically, I cannot
rigorously rule out some sort of chaotic behavior in the phases of the fields,
yet, but since the behavior of all the field bilinears is so deterministic,
including those involving spatial derivatives, my current opinion is that
chaotic behavior in the fields, if any, must be limited to only time-dependent
phases and is completely innocuous. \ (I would devote more time to this issue,
but once again, I view this model as just a warm-up exercise for the $A^{2}$
model presented below.) \ I leave it to the reader to compare the above system
of equations to those of the standard isotropic oscillator \cite{Lewis,Mahan}.

For consistency with the treatment of the point-particle case in \cite{CPZ},
and to warm the hearts of conformal field theorists, the above analysis should
perhaps be re-done using complex coordinates. \ Also, a generalization to a
polygon area potential, involving $N$ types of particles and $A\left(
\mathbf{r}_{1},\mathbf{r}_{2},\cdots,\mathbf{r}_{N}\right)  =\mathbf{r}%
_{1}\wedge\mathbf{r}_{2}+\mathbf{r}_{2}\wedge\mathbf{r}_{3}+\cdots
+\mathbf{r}_{N}\wedge\mathbf{r}_{1}$, is straightforward. \ An elegant touch
would be to introduce a tensor $E^{a_{1}\cdots a_{N}}$ on the particle type
such that the combination $A\left(  \mathbf{r}_{1},\mathbf{r}_{2}%
,\cdots,\mathbf{r}_{N}\right)  E^{a_{1}\cdots a_{N}}$ is symmetric under any
pair interchange $\mathbf{r}_{i},a_{i}\leftrightarrow\mathbf{r}_{j},a_{j}$.
\ Again the most streamlined way to do this on the plane is to work with
complex coordinates, as was the case in the point-particle situation.

We next construct an area-squared field theory, where the potential is $A^{2}%
$, a quartic function of field coordinates. \ This can be incorporated into a
many-body field theory with only a single complex field $\psi$ in any number
of spatial dimensions $n$. \ The Hamiltonian for the model is%
\begin{align*}
H[\psi]  &  =\frac{1}{2m}\int d^{n}s\,\partial_{i}\bar{\psi}\partial_{i}\psi\\
&  +\tfrac{1}{3}k\int d^{n}s_{1}\int d^{n}s_{2}\int d^{n}s_{3}\,\left|
\psi\left(  s_{1}\right)  \right|  ^{2}\left|  \psi\left(  s_{2}\right)
\right|  ^{2}\left|  \psi\left(  s_{3}\right)  \right|  ^{2}\left[
s_{{\footnotesize 12}}^{2}s_{{\footnotesize 23}}^{2}-\left(
s_{{\footnotesize 12}}\cdot s_{{\footnotesize 23}}\right)  ^{2}\right]
\end{align*}
where $\mathbf{s}_{ab}\equiv\mathbf{s}_{a}-\mathbf{s}_{b}$ for each $ab$ pair
of the identical particles' coordinate vectors. \ The equations of motion from
$L\left[  \psi\right]  =\int d^{n}s\,i\bar{\psi}\partial_{t}\psi-H[\psi]$ are%
\[
i\partial_{t}\psi\left(  r\right)  +\frac{1}{2m}\nabla^{2}\psi\left(
r\right)  =V\left[  \psi\right]  \left(  r\right)  \,\psi\left(  r\right)
\]
In this case the effective potential corresponds to a non-isotropic
oscillator
\begin{align*}
V\left[  \psi\right]  \left(  r\right)   &  =k\int d^{n}s_{2}\int d^{n}%
s_{3}\,\left|  \psi\left(  s_{2}\right)  \right|  ^{2}\left|  \psi\left(
s_{3}\right)  \right|  ^{2}\left.  \left[  s_{{\footnotesize 12}}%
^{2}s_{{\footnotesize 23}}^{2}-\left(  s_{{\footnotesize 12}}\cdot
s_{{\footnotesize 23}}\right)  ^{2}\right]  \right|  _{s_{1}=r}\\
&  =K_{ij}r^{i}r^{j}+B_{i}r^{i}+Z=K_{ij}\left(  r^{i}r^{j}-2R^{i}r^{j}\right)
+Z\\
&  =K_{ij}\left(  r^{i}-R^{i}\right)  \left(  r^{j}-R^{j}\right)
+Z-K_{ij}R^{i}R^{j}%
\end{align*}
with the definitions and relations
\begin{align*}
K_{ij}  &  \equiv k\int d^{n}s_{2}\int d^{n}s_{3}\,\left|  \psi\left(
s_{2}\right)  \right|  ^{2}\left|  \psi\left(  s_{3}\right)  \right|
^{2}\left(  \delta^{ij}s_{23}^{2}-s_{23}^{i}s_{23}^{j}\right) \\
&  =2kN^{2}\left(  \left(  S-R^{2}\right)  \delta_{ij}-\left(  S_{ij}%
-R_{i}R_{j}\right)  \right)
\end{align*}%
\begin{align*}
B_{i}  &  \equiv-2k\int d^{n}s_{2}\int d^{n}s_{3}\,\left|  \psi\left(
s_{2}\right)  \right|  ^{2}\left|  \psi\left(  s_{3}\right)  \right|
^{2}\left(  s_{{\footnotesize 2}}^{i}\,s_{{\footnotesize 23}}^{2}%
-s_{{\footnotesize 23}}^{i}\,s_{{\footnotesize 2}}\cdot s_{{\footnotesize 23}%
}\right) \\
&  =4kN^{2}\left(  S_{ij}R_{j}-SR_{i}\right)  =-2K_{ij}R_{j}%
\end{align*}%
\[
Z\equiv k\int d^{n}s_{2}\int d^{n}s_{3}\,\left|  \psi\left(  s_{2}\right)
\right|  ^{2}\left|  \psi\left(  s_{3}\right)  \right|  ^{2}\left(  s_{2}%
^{2}s_{3}^{2}-\left(  s_{2}\cdot s_{3}\right)  ^{2}\right)  =kN^{2}\left(
S^{2}-S_{ij}S_{ji}\right)
\]%
\begin{align*}
N  &  \equiv%
{\textstyle\int}
d^{n}s\,\left|  \psi\left(  s\right)  \right|  ^{2}\;,\;\;\;R_{i}%
\equiv\frac{1}{N}%
{\textstyle\int}
d^{n}s\,s_{i}\left|  \psi\left(  s\right)  \right|  ^{2}\;,\;\;\;P_{i}\left[
\psi\right]  \equiv\frac{-i}{N}%
{\textstyle\int}
d^{n}s\left(  \bar{\psi}\overleftrightarrow{\partial_{i}}\psi\right)  \;,\\
S_{ij}  &  \equiv\frac{1}{N}%
{\textstyle\int}
d^{n}s\,s_{i}s_{j}\left|  \psi\left(  s\right)  \right|  ^{2}\;,\;\;\;S=\sum
_{j=1}^{n}S_{jj}\;,\;\;\;K_{ij}=2kN^{2}\left(  S\delta_{ij}-S_{ij}\right)  \;.
\end{align*}
Note that everything in $V\left[  \psi\right]  $ factorizes into integrals of
bilinear densities. \ Actually, this is not too surprising given the
polynomial character and factorized form of the terms of the three-body
potential. \ So the effective potential for the area$^{2}$ model becomes
\begin{align*}
V\left[  \psi\right]  \left(  r\right)   &  =2kN^{2}\left[  S\left(
r-R\right)  ^{2}-S_{ij}\left(  r^{i}-R^{i}\right)  \left(  r^{j}-R^{j}\right)
\right]  +2kN^{2}\left(  S_{ij}R^{i}R^{j}-SR^{2}\right) \\
&  +kN^{2}\left(  S^{2}-S_{ij}S_{ij}\right)  -2kN^{2}\left(  R^{2}%
r^{2}-\left(  r\cdot R\right)  ^{2}\right)
\end{align*}
As already noted, this is a non-isotropic quadratic potential. \ Note the
presence of the last term $R^{2}r^{2}-\left(  r\cdot R\right)  ^{2}=\left(
\mathbf{r}\wedge\mathbf{R}\right)  ^{2}$ which is the squared area of the
parallelogram formed from $r$ and $R$. \ The negative coefficient of this term
and the potential instability it would produce for large $r\perp R$ are
completely offset by the additional presence of $2kN^{2}Sr^{2}$ in the
effective potential, along with the standard inequality $S\geq R^{2}$. \ Also
note that $V$ is non-zero only through the field spreading out around the
center of mass location. \ Were $\left|  \psi\left(  s\right)  \right|  ^{2}$
a delta-function at $R$, all coefficients would vanish in $V$.

Also note the absence of \emph{nontrivial} ``zero modes'' for the quadratic
part of $V$ (i.e. those zero-modes that would persist even in the CM frame,
and not those associated with the free translation of the CM). \ This is
straightforward to see, even though for the field coordinates alone
$\det\left(  s_{i}s_{j}\right)  =0=\det\left(  s^{2}\delta_{ij}-s_{i}%
s_{j}\right)  $,\ and so these coordinate matrices have zero eigenvalues.
\ However, $\left\langle \det S\right\rangle \neq\det\left\langle
S\right\rangle $ just as $\left\langle x^{2}\right\rangle \neq\left\langle
x\right\rangle ^{2}$. \ In fact $\det\left(  S\delta_{ij}-S_{ij}\right)
=\frac{1}{N^{n}}\det\int d^{n}s\,\left(  s^{2}\delta_{ij}-s_{i}s_{j}\right)
\,\left|  \psi\left(  s\right)  \right|  ^{2}\geq0$. \ Moreover, in the CM
$\det K>0$ \emph{and all the eigenvalues of }$K_{ij}$\emph{\ are greater than
zero}. \ Or equivalently, all the eigenvalues of $S\delta_{ij}-S_{ij}$ are
positive. \ Proof: \ For any real vector $v$, we have $v_{i}\left(
S\delta_{ij}-S_{ij}\right)  v_{j}=\frac{1}{N}\int d^{n}s\,\left(  s^{2}%
v^{2}-\left(  s\cdot v\right)  ^{2}\right)  \,\left|  \psi\left(  s\right)
\right|  ^{2}$. \ But the integrand is non-negative by the Schwarz inequality
$s^{2}v^{2}-\left(  s\cdot v\right)  ^{2}\geq0$. \ Thus any eigenvalue
$\lambda$ of $K_{ij}$ is non-negative, $\lambda\geq0$. \ Now to rule out the
case where any eigenvalue vanishes in the CM, we note the Schwarz inequality
is only an equality $s^{2}v^{2}-\left(  s\cdot v\right)  ^{2}=0$ if
$s_{i}\propto v_{i}$. \ Therefore the only way to have a zero eigenvalue is to
have the full integrand $\left(  s^{2}v^{2}-\left(  s\cdot v\right)
^{2}\right)  \,\left|  \psi\left(  s\right)  \right|  ^{2}$ restricted to a
line where $s_{i}\propto v_{i}$. \ Forget that! \ There is always some
transverse spread in $\left|  \psi\left(  s\right)  \right|  ^{2}$. \ Any
reasonable wave function (or classical field) in $n\geq2$ dimensions does not
have support just on a line. \ So $V\left[  \psi\right]  \left(  r\right)  $
always has a quadratic $r^{2}$ part and therefore the field is always
localized around the CM. Or, perhaps more physically, all the particles are
bound by the effective potential to the CM.

For solutions the energy is
\[
H[\psi]=\frac{1}{8m}NT+kN^{3}\left(  \mathcal{S}^{2}-\mathcal{S}%
_{ij}\mathcal{S}_{ij}\right)
\]
using $\mathcal{S}_{ij}=S_{ij}-R_{i}R_{j}\;,\;\;\mathcal{S}=\mathcal{S}%
_{jj}\;,\;\;T=T_{jj}\;,$ with the stress-tensor defined as%
\[
T_{ij}\left[  \psi\right]  \equiv\frac{-1}{N}\int d^{n}s\,\left(  \bar{\psi
}\overleftrightarrow{\partial_{i}}\overleftrightarrow{\partial_{j}}%
\psi\right)  =\frac{-4}{N}\int d^{n}s\,\left(  \bar{\psi}\partial_{i}%
\partial_{j}\psi\right)
\]

Now what is the time-dependence of the three effective potential coefficients
$K_{ij},\;B_{i},\;$and $Z$? \ Or rather, what is the time-dependence of the
quantities that compose them? \ Obviously, $N,$ $R_{i},$ and $P_{i}$ obey the
equations
\[
\frac{d}{dt}N=0\;,\;\;\;\;\;\frac{d}{dt}P_{i}=0\;,\;\;\;\;\;\frac{d}{dt}%
R_{i}=\frac{1}{2m}\,P_{i}%
\]
with the solutions
\[
N\left(  t\right)  =N_{0}\equiv N\;,\;\;\;\;\;P_{i}\left(  t\right)
=P_{0i}\equiv P_{i}\;,\;\;\;\;\;R_{i}\left(  t\right)  =R_{0i}+\frac{1}%
{2m}\,P_{i}t
\]
But what about $S_{ij}$? \ With the definition (note that $Q_{ij}\neq Q_{ji}%
$)
\[
Q_{ij}\left[  \psi\right]  \equiv\frac{-i}{N}\int d^{n}s\,s_{i}\,\left(
\bar{\psi}\overleftrightarrow{\partial_{j}}\psi\right)  =\frac{-2i}{N}\int
d^{n}s\,s_{i}\,\left(  \bar{\psi}\partial_{j}\psi\right)  -i\delta_{ij}%
\]
we find
\[
\frac{d}{dt}S_{ij}=\frac{1}{2m}\,\left(  Q_{ij}+Q_{ji}\right)
\]
So we must include $Q_{ij}$ into the mix of differential equations. \ It's
time derivative is
\[
\frac{d}{dt}Q_{ij}=\frac{1}{2m}\,T_{ij}-4S_{ik}K_{kj}-2R_{i}B_{j}%
\]
where $T_{ij}\left[  \psi\right]  $ is the previous stress-tensor. \ We are
one step away from closing the system. \ We just need the final time
derivative
\[
\frac{d}{dt}T_{ij}=-4\left(  K_{ik}Q_{kj}+K_{jk}Q_{ki}\right)  -2P_{i}%
B_{j}-2P_{j}B_{i}%
\]
The system of equations is now closed. \ The problem then is to solve these
equations to obtain $S_{ij}$ and hence the effective potential in terms of the
initial field configuration . \ Note that only the symmetric tensor part of
$Q_{ij}$ varies with time. \ 

Let's break down the tensors into somewhat more traditional combinations.
\ The symmetric and antisymmetric parts of $Q_{ij}$ are (almost) familiar from
rotations and (at least when traced) coordinate rescaling. \
\[
J_{ij}=Q_{ij}-Q_{ji}\;,\;\;\;\;\;D_{ij}=Q_{ij}+Q_{ji}%
\]
These nicely occur above with their values for (or about) the CM removed.
\ Thus define
\[
\mathcal{D}_{ij}=D_{ij}-\left(  R_{i}P_{j}+R_{j}P_{i}\right)
\;,\;\;\;\;\;\mathcal{J}_{ij}=J_{ij}-\left(  R_{i}P_{j}-R_{j}P_{i}\right)
\]
Similarly remove from the quadrupole and stress tensors the CM coordinate and
momentum dyads.
\[
\mathcal{S}_{ij}=S_{ij}-R_{i}R_{j}\;,\;\;\;\;\;\mathcal{T}_{ij}=T_{ij}%
-P_{i}P_{j}%
\]
Incorporating momentum and angular momentum conservation, i.e. $B_{i}%
=-2K_{ij}R^{j}$ and $K_{ik}S_{kj}-R_{j}K_{ik}R^{k}=K_{jk}S_{ki}-R_{i}%
K_{jk}R^{k}$ (just a matrix commutator,$\;\left[  \mathcal{S},K\right]
_{ij}=0$), the time evolution equations are then
\begin{align}
\frac{d}{dt}\mathcal{S}_{ij} &  =\frac{1}{2m}\,\mathcal{D}_{ij}%
\;,\;\;\;\;\;\frac{d}{dt}N=0\;,\;\;\;\;\;\frac{d}{dt}\mathcal{J}%
_{ij}=0\label{AnisotropicEqns}\\
\frac{d}{dt}\mathcal{D}_{ij} &  =\frac{1}{m}\,\mathcal{T}_{ij}-4\mathcal{S}%
_{ik}K_{kj}-4\mathcal{S}_{jk}K_{ki}=\frac{1}{m}\,\mathcal{T}_{ij}-4\left\{
\mathcal{S},K\right\}  _{ij}\nonumber\\
\frac{d}{dt}\mathcal{T}_{ij} &  =-2K_{ik}\left(  \mathcal{D}_{kj}%
+\mathcal{J}_{kj}\right)  -2K_{jk}\left(  \mathcal{D}_{ki}+\mathcal{J}%
_{ki}\right)  =-2\left\{  K,\mathcal{D}\right\}  _{ij}-2\left[  K,\mathcal{J}%
\right]  _{ij}\nonumber
\end{align}
where now $K_{ij}=2kN^{2}\left(  \mathcal{S}\delta_{ij}-\mathcal{S}%
_{ij}\right)  \;$. \ Anyone skilled in the black arts of nuclear physics
should feel comfortable with equations of this sort.

An invariant arises from taking traces. \ Thus%
\[
\frac{d}{dt}\left(  \operatorname*{tr}\mathcal{T}_{ij}+8mkN^{2}\left(
\mathcal{S}^{2}-\operatorname*{tr}\mathcal{S}_{ij}^{2}\right)  \right)  =0\;.
\]
Now, this is essentially just the energy in the center-of-mass.%
\[
P^{2}+\operatorname*{tr}\mathcal{T}_{ij}+8mkN^{2}\left(  \mathcal{S}%
^{2}-\operatorname*{tr}\mathcal{S}_{ij}^{2}\right)  =\operatorname*{tr}%
T_{ij}+8mkN^{2}\left(  \mathcal{S}^{2}-\operatorname*{tr}\mathcal{S}_{ij}%
^{2}\right)  =8mH[\psi]/N
\]
where%
\[
H[\psi]=\frac{1}{8m}NT+kN^{3}\left(  \mathcal{S}^{2}-\mathcal{S}%
_{ij}\mathcal{S}_{ij}\right)  \;.
\]
As a special situation, but really without loss of generality, consider
co-moving with the system in the CM, where $R_{i}=0=P_{i}$. \ Then the
effective potential coefficients reduce to
\[
K_{ij}=2kN^{2}\left(  \mathcal{S}\delta_{ij}-\mathcal{S}_{ij}\right)
\;,\;\;\;\;\;B_{i}=0\;,\;\;\;\;\;Z=kN^{2}\left(  \mathcal{S}^{2}%
-\mathcal{S}_{ij}\mathcal{S}_{ij}\right)
\]
with the $\mathcal{S}_{ij},\;\mathcal{D}_{ij},$ and $\mathcal{T}_{ij}$ time
derivatives unchanged in form. \ (Note in the CM $S_{ij}=\mathcal{S}%
_{ij},\;D_{ij}=\mathcal{D}_{ij},$ and $T_{ij}=\mathcal{T}_{ij}$.) \ 

For now, I cannot solve the non-isotropic equations (\ref{AnisotropicEqns}).
\ Perhaps closed-form solutions can be obtained in various limiting
situations, but most probably the system is chaotic for general initial
data\footnote{Since giving this talk, extensive numerical investigations have
shown that the model is indeed chaotic for general initial data \cite{CK}.}.
\ Nevertheless, I can solve exactly a further specialization, which \emph{is}
with loss of generality. \ Suppose the various densities are actually
isotropic about the CM. \ Then we have
\begin{align*}
\mathcal{D}_{ij}  &  =\frac{1}{n}\mathcal{D}\delta_{ij}\;,\;\;\;\mathcal{S}%
_{ij}=\frac{1}{n}\mathcal{S}\delta_{ij}\;,\;\;\;\mathcal{T}_{ij}=\frac{1}%
{n}\mathcal{T}\delta_{ij}\;,\;\;\;K_{ij}=\frac{1}{n}K\delta_{ij}%
=2kN^{2}\left(  \frac{n-1}{n}\right)  \mathcal{S}\delta_{ij}\\
K  &  =2kN^{2}\left(  n-1\right)  \mathcal{S}\;,\;\;\;Z=kN^{2}\left(
\frac{n-1}{n}\right)  \mathcal{S}^{2}%
\end{align*}
and the time derivatives are $\frac{d}{dt}K_{ij}=2kN^{2}\left(  \frac{n-1}%
{n}\right)  \frac{1}{2m}\,\mathcal{D}\delta_{ij}$, etc. \ These may all be
traced now w.l.o.g. to yield
\[
\frac{d}{dt}K=\frac{1}{m}\,kN^{2}\left(  n-1\right)  \mathcal{D}%
\;,\;\;\;\;\;\frac{d}{dt}Z=\frac{1}{m}\,kN^{2}\left(  \frac{n-1}{n}\right)
\mathcal{SD}\;,\;\;\;\;\;\frac{d}{dt}\mathcal{S}=\frac{1}{2m}\,\mathcal{D}\;,
\]
and the remaining equations%
\[
\frac{d}{dt}\mathcal{D}=\frac{1}{m}\,\mathcal{T}-16kN^{2}\left(  \frac{n-1}%
{n}\right)  \mathcal{S}^{2}\;,\;\;\;\frac{d}{dt}\mathcal{T}=-8kN^{2}\left(
\frac{n-1}{n}\right)  \mathcal{SD}%
\]
Using $\frac{d}{dt}\mathcal{S}=\frac{1}{2m}\,\mathcal{D}$ the $\mathcal{T}$
equation becomes $\frac{d}{dt}\mathcal{T}=-8mkN^{2}\left(  \frac{n-1}%
{n}\right)  \frac{d}{dt}\mathcal{S}^{2}$ which integrates immediately to
\[
\mathcal{T}\left(  t\right)  =\mathcal{T}_{0}+8mkN^{2}\left(  \frac{n-1}%
{n}\right)  \left(  \mathcal{S}_{0}^{2}-\mathcal{S}\left(  t\right)
^{2}\right)
\]
Inserting this into the $\mathcal{D}$ equation and using $\frac{d}%
{dt}\mathcal{S}=\frac{1}{2m}\,\mathcal{D}$ once again, we have a second-order
equation for $\mathcal{S}$.%
\[
2m\,\frac{d^{2}}{dt^{2}}\mathcal{S}=\frac{1}{m}\,\mathcal{T}_{0}%
+8kN^{2}\left(  \frac{n-1}{n}\right)  \mathcal{S}_{0}^{2}-24kN^{2}\left(
\frac{n-1}{n}\right)  \mathcal{S}^{2}%
\]
Thus $\mathcal{S}$ alone evolves as a nonlinear oscillator subject to a cubic
self-interaction. \ That is, the effective potential governing the time
evolution of $\mathcal{S}$ is $\mathcal{V}\left(  \mathcal{S}\right)
=a\mathcal{S}^{3}-b\mathcal{S}$, \ where the constant, initial-data-dependent
coefficients are $a=8kN^{2}\left(  \frac{n-1}{n}\right)  \;,$ $b=\frac{1}%
{m}\,\mathcal{T}_{0}+8kN^{2}\left(  \frac{n-1}{n}\right)  \mathcal{S}_{0}^{2}%
$, and are both positive. \ So the potential has a local minimum/maximum at
$3a\mathcal{S}^{2}=b$ or $\mathcal{S}_{\min/\max}=\pm\sqrt{\frac{b}{3a}}$,
hence
\[
\mathcal{S}_{\min/\max}=\pm\sqrt{\frac{1}{3}\mathcal{S}_{0}^{2}+\frac{1}%
{24mkN^{2}}\left(  \frac{n}{n-1}\right)  \mathcal{T}_{0}}%
\]
Only the local minimum here is physical. \ Recall that we are in the CM, so
that $D=\mathcal{D}$, $T=\mathcal{T}$, and most importantly the inequality
$\mathcal{S}=S\equiv\frac{1}{N}\int d^{n}s\,s^{2}\,\left|  \psi\left(
s\right)  \right|  ^{2}\geq0\;,$so we see that physically $\mathcal{S}$ is
constrained to be positive (and $\mathcal{S}=0$ only for $\left|  \psi\left(
s\right)  \right|  ^{2}$ ultra-localized at the origin, like a delta-function
$\cdots$ a collapse/condensation of the matter field).

Is it possible that we encounter singularities here, in a finite time? \ By
it's definition we are constrained to have $S\geq0$,\ but for ``special''
initial data, perhaps the equation evolves $S$ to zero, the boundary of the
unphysical $S<0$ region, hence the matter field would collapse to the origin.
\ After all, the equation for $S$ is that of a nonlinear ``cubic'' oscillator.
\ Do we have here a baby version of black-hole physics?\ 

The answer is no, $S=0$ is never achieved, as we now explain. \ The effective
potential for $\mathcal{S}$ will require $\mathcal{S}>0$ to be true \emph{for
all times} if and only if $\mathcal{S}_{0}>0$ and $\mathcal{E}\left(
\mathcal{S}\right)  <0$, where $\mathcal{E}$ is the conserved \emph{effective
energy} for $\mathcal{S}$.
\[
\mathcal{E}\left(  \mathcal{S}\right)  \equiv m\left(  \frac{d}{dt}%
\mathcal{S}\right)  ^{2}+\mathcal{V}\left(  \mathcal{S}\right)  =\mathcal{E}%
_{0}%
\]%
\[
\mathcal{E}_{0}=\frac{1}{4m}\mathcal{D}_{0}^{2}+8kN^{2}\left(  \tfrac{n-1}%
{n}\right)  \mathcal{S}_{0}^{3}-\left(  \frac{1}{m}\,\mathcal{T}_{0}%
+8kN^{2}\left(  \tfrac{n-1}{n}\right)  \mathcal{S}_{0}^{2}\right)
\mathcal{S}_{0}=\frac{1}{4m}\left(  \mathcal{D}_{0}^{2}-4\,\mathcal{T}%
_{0}\mathcal{S}_{0}\right)
\]
Consider the two terms on the RHS of this last expression. \ The standard QM
uncertainty relation (basically the Schwarz inequality) for any two hermitean
operators $\mathbf{a}$ and $\mathbf{b}$, assuming $\left\langle \mathbf{a}%
\right\rangle =0=\left\langle \mathbf{b}\right\rangle $, is $\left\langle
\mathbf{a}^{2}\right\rangle \left\langle \mathbf{b}^{2}\right\rangle
\geq\left\langle \frac{-i}{2}\left(  \mathbf{ab}-\mathbf{ba}\right)
\right\rangle ^{2}+\left\langle \frac{1}{2}\left(  \mathbf{ab}+\mathbf{ba}%
\right)  \right\rangle ^{2}$. \ Each of the two terms on the RHS of the
inequality are positive. \ Usually we discard the second of these RHS terms
and keep the first, especially in the case for canonically conjugate variables
with $\mathbf{xp}-\mathbf{px}=i\hbar$. \ But we could equally well discard the
first RHS term to obtain$\left\langle \mathbf{a}^{2}\right\rangle \left\langle
\mathbf{b}^{2}\right\rangle -\left\langle \frac{1}{2}\left(  \mathbf{ab}%
+\mathbf{ba}\right)  \right\rangle ^{2}\geq0$. \ This is a strict inequality
for conjugate variables. \ Applying this to the situation at hand we conclude
$4\mathcal{TS}-\mathcal{D}^{2}>0$. \ Hence the cubic oscillator system above
is always bound to the stable minimum, with $\mathcal{E}<0$. \ This is what we
wanted to show.

We will discuss static solutions in more detail below. \ For now, we note that
these are possible if we are at the local minimum of $\mathcal{V}\left(
\mathcal{S}\right)  $ with $\mathcal{D}=0$, where we require
\[
\mathcal{S}_{\min}=\mathcal{S}_{0}=\sqrt{\frac{1}{3}\mathcal{S}_{0}%
^{2}+\frac{1}{24mkN^{2}}\left(  \frac{n}{n-1}\right)  \mathcal{T}_{0}%
}\;,\;\;\;\text{or}\;\;\;\mathcal{T}_{0}=16mkN^{2}\left(  \frac{n-1}%
{n}\right)  \mathcal{S}_{0}^{2}\;,
\]
as well as $\mathcal{D}_{0}=0$, of course. \ The fields themselves may be
worked out explicitly in this case and are nothing but Gaussians.

Even in the non-static case the conserved energy for $\mathcal{S}$ reduces
it's determination to quadrature and allows us to solve for the $\mathcal{S}$
motion in textbook fashion.%
\[
\frac{d}{dt}\mathcal{S=\pm}\sqrt{\left(  \mathcal{E}_{0}-\mathcal{V}\left(
\mathcal{S}\right)  \right)  /m}%
\]%
\[
\pm\frac{t}{\sqrt{m}}=\int_{\mathcal{S}_{0}}^{\mathcal{S}\left(  t\right)
}\frac{d\mathcal{S}}{\sqrt{\mathcal{E}_{0}-\mathcal{V}\left(  \mathcal{S}%
\right)  }}=\frac{1}{\sqrt{a}}\int_{\mathcal{S}_{0}}^{\mathcal{S}\left(
t\right)  }\frac{d\mathcal{S}}{\sqrt{\left(  \mathcal{S}_{hi}-\mathcal{S}%
\right)  \left(  \mathcal{S}-\mathcal{S}_{mid}\right)  \left(  \mathcal{S}%
-\mathcal{S}_{lo}\right)  }}%
\]
where for $\mathcal{E}_{0}<0$, $\left\{  \mathcal{S}_{lo},\mathcal{S}%
_{mid},\mathcal{S}_{hi}\right\}  $ are the three distinct zeroes of the cubic
$\mathcal{E}_{0}+b\mathcal{S}-a\mathcal{S}^{3}$\ with $\mathcal{S}%
_{lo}<0<\mathcal{S}_{mid}\leq\mathcal{S}\left(  t\right)  \leq\mathcal{S}%
_{hi}$. \ The RHS here is an elliptic integral.%
\[
\int_{\mathcal{S}_{mid}}^{\mathcal{S}\left(  t\right)  }\tfrac{\sqrt
{\mathcal{S}_{hi}-\mathcal{S}_{lo}}\;\;dS\;\;}{\sqrt{4\left(  \mathcal{S}%
_{hi}-S\right)  \left(  S-\mathcal{S}_{mid}\right)  \left(  S-\mathcal{S}%
_{lo}\right)  }}=K\left(  \sqrt{\tfrac{\mathcal{S}_{hi}-\mathcal{S}_{mid}%
}{\mathcal{S}_{hi}-\mathcal{S}_{lo}}}\right)  -F\left(  \sqrt{\tfrac
{\mathcal{S}_{hi}-\mathcal{S}\left(  t\right)  }{\mathcal{S}_{hi}%
-\mathcal{S}_{mid}}},\sqrt{\tfrac{\mathcal{S}_{hi}-\mathcal{S}_{mid}%
}{\mathcal{S}_{hi}-\mathcal{S}_{lo}}}\right)
\]
$K$ and $F$ are the standard elliptic integrals, with\textbf{ }$F\left(
1,z\right)  =K\left(  z\right)  $, and $F\left(  0,z\right)  =0$. \ Thus
$\mathcal{S}\left(  t\right)  $ is an elliptic function. \ The solution
oscillates between the distinct turning points $\mathcal{S}_{mid}%
\leq\mathcal{S}\left(  t\right)  \leq\mathcal{S}_{hi}$ with period%
\[
\tfrac{1}{2}t_{period}=\sqrt{\tfrac{m}{a}}\,\int_{\mathcal{S}_{mid}%
}^{\mathcal{S}_{hi}}\frac{d\mathcal{S}}{\sqrt{\left(  \mathcal{S}%
_{hi}-\mathcal{S}\right)  \left(  \mathcal{S}-\mathcal{S}_{mid}\right)
\left(  \mathcal{S}-\mathcal{S}_{lo}\right)  }}=\sqrt{\tfrac{4m}{a\left(
\mathcal{S}_{hi}-\mathcal{S}_{lo}\right)  }}K\left(  \sqrt{\tfrac
{\mathcal{S}_{hi}-\mathcal{S}_{mid}}{\mathcal{S}_{hi}-\mathcal{S}_{lo}}%
}\right)
\]

We may extend the analysis to higher dimensions for models with potentials
depending on higher multi-particle coordinate forms. \ Assume a ($d+1$)-body
potential $U$ in $n\geq d$ dimensions which is a ($d$-form)$^{2}$. \ That is
\[
F\left(  \mathbf{r}_{1},\mathbf{r}_{2},\mathbf{r}_{3},\cdots,\mathbf{r}%
_{d},\mathbf{r}_{d+1}\right)  =(\mathbf{r}_{1}-\mathbf{r}_{2})\wedge
(\mathbf{r}_{2}-\mathbf{r}_{3})\wedge\cdots\wedge(\mathbf{r}_{d}%
-\mathbf{r}_{d+1})
\]%
\begin{align*}
U\left(  \mathbf{r}_{1},\mathbf{r}_{2},\cdots,\mathbf{r}_{d+1}\right)   &
=kF^{2}\left(  \mathbf{r}_{1},\mathbf{r}_{2},\cdots,\mathbf{r}_{d+1}\right) \\
&  =k\;\delta_{j_{1}j_{2}\cdots j_{d}}^{i_{1}i_{2}\cdots i_{d}}\;\mathbf{r}%
_{12}^{i_{1}}\mathbf{r}_{12}^{j_{1}}\mathbf{r}_{23}^{i_{2}}\mathbf{r}%
_{23}^{j_{2}}\mathbf{r}_{34}^{i_{3}}\mathbf{r}_{34}^{j_{3}}\cdots
\mathbf{r}_{dd+1}^{i_{d}}\mathbf{r}_{dd+1}^{j_{d}}%
\end{align*}%
\begin{align*}
H  &  =\int\left(  d\mathbf{r}\right)  \;\psi^{\ast}\left(  \mathbf{r}\right)
\left[  -\frac{1}{2m}\nabla^{2}\right]  \psi\left(  \mathbf{r}\right) \\
&  +\frac{1}{d+1}\idotsint\left(  d\mathbf{r}_{1}\right)  \cdots\left(
d\mathbf{r}_{d+1}\right)  \;U\left(  \mathbf{r}_{1},\mathbf{r}_{2}%
,\cdots,\mathbf{r}_{d+1}\right)  \;\left|  \psi\left(  \mathbf{r}%
_{d+1}\right)  \cdots\psi\left(  \mathbf{r}_{2}\right)  \psi\left(
\mathbf{r}_{1}\right)  \right|  ^{2}%
\end{align*}
with $\idotsint\left(  d\mathbf{r}_{2}\right)  \cdots\left(  d\mathbf{r}%
_{d+1}\right)  \;U\left(  \mathbf{r},\mathbf{r}_{2},\cdots,\mathbf{r}%
_{d+1}\right)  \;\left|  \psi\left(  \mathbf{r}_{d+1}\right)  \cdots
\psi\left(  \mathbf{r}_{2}\right)  \right|  ^{2}\;\psi\left(  \mathbf{r}%
\right)  $ in the field equation.

If we assume spherically symmetric fields we always obtain from this isotropic
harmonic effective potentials.
\begin{align*}
&  \idotsint\left(  d\mathbf{r}_{2}\right)  \cdots\left(  d\mathbf{r}%
_{d+1}\right)  \;U\left(  \mathbf{r},\mathbf{r}_{2},\cdots,\mathbf{r}%
_{d+1}\right)  \;\left|  \psi\left(  \mathbf{r}_{d+1}\right)  \cdots
\psi\left(  \mathbf{r}_{2}\right)  \right|  ^{2}\\
&  =k\left(  n-1\right)  \left(  n-2\right)  \cdots\left(  n-d+1\right)
\left(  dM_{0}r^{2}+M_{2}\right)  \left(  \frac{M_{2}}{n}\right)  ^{d-1}%
\end{align*}
where $M_{k}\equiv\int\left(  d\mathbf{r}\right)  \left(  \mathbf{r}%
^{2}\right)  ^{k/2}\left|  \psi\left(  r\right)  \right|  ^{2}=\Omega_{n}%
\int_{0}^{\infty}\left|  \psi\right|  ^{2}r^{k+n-1}dr$ and $\Omega_{n}%
=2\pi^{n/2}/\Gamma\left(  n/2\right)  $. \ As usual the interpretation (at
least in the quantum theory) is that $N\equiv M_{0}$ is the total number of
particles. \ Thus the field equation for such spherically symmetric solutions
becomes
\[
i\frac{\partial}{\partial t}\psi\left(  r\right)  +\frac{1}{2m}\left(
\frac{d^{2}}{dr^{2}}\psi\left(  r\right)  +\frac{n-1}{r}\frac{d}{dr}%
\psi\left(  r\right)  \right)  =\frac{M_{2}^{d-1}\left(  n-1\right)
!}{n^{d-1}\left(  n-d\right)  !}\left(  dM_{0}r^{2}+M_{2}\right)  k\psi\left(
r\right)  \;.
\]
For a Gaussian ansatz, $\psi\left(  \mathbf{r},t\right)  =C\exp\left(
-\frac{1}{2}c\mathbf{r}^{2}\right)  \exp\left(  -iEt\right)  $, this field
equation reduces to
\[
E+\frac{1}{2m}\left(  c^{2}r^{2}-nc\right)  =k\frac{\left(  n-1\right)
!}{\left(  n-d\right)  !}\left(  \frac{N}{2c}\right)  ^{d-1}\left(
dNr^{2}+\frac{nN}{2c}\right)
\]
and so we have a solution provided
\[
\frac{1}{2m}c^{d+1}=k\frac{\left(  n-1\right)  !}{\left(  n-d\right)  !}%
N^{d}\frac{d}{2^{d-1}}\;,\;\;\;\;\;E=\frac{nc}{2m}\left(  1+\frac{1}%
{2d}\right)
\]

The $E$ above is almost, but not quite, the energy per particle. \ That is,
the value of $H$, the total energy, is\ almost but not quite $EN$ for the
spherically symmetric solution. \ For the Gaussian ansatz%
\[
H=\frac{ncN}{4m}\left(  1+\frac{1}{d}\right)
\]

The above suggests what to do also in the situation where $\psi$ is not
necessarily spherically symmetric. \ For any $U\left(  \mathbf{r}%
,\mathbf{r}_{2},\mathbf{r}_{3}\cdots,\mathbf{r}_{d},\mathbf{r}_{d+1}\right)  $
which is a quadratic form in the components of $\mathbf{r}$ we may write quite
generally
\[
Z+\mathbf{r}^{i}B_{i}+\mathbf{r}^{i}\mathbf{r}^{j}K_{ij}=%
{\textstyle\idotsint}
\left(  d\mathbf{r}_{2}\right)  \cdots\left(  d\mathbf{r}_{d+1}\right)
\;U\left(  \mathbf{r},\mathbf{r}_{2},\cdots,\mathbf{r}_{d+1}\right)  \;\left|
\psi\left(  \mathbf{r}_{d+1}\right)  \cdots\psi\left(  \mathbf{r}_{2}\right)
\right|  ^{2}\;.
\]
This is true for the case at hand. \ $U\left(  \mathbf{r},\mathbf{r}%
_{2},\mathbf{r}_{3}\cdots,\mathbf{r}_{d},\mathbf{r}_{d+1}\right)  $ is a
quadratic form in the components of $\mathbf{r}$ (as well as quadratic in all
the other individual $\mathbf{r}_{a}$) even though it is of order $2d$
altogether. \ Hence
\[
K_{ij}=k\delta_{jj_{2}\cdots j_{d}}^{ii_{2}\cdots i_{d}}\,%
{\textstyle\idotsint}
\left(  d\mathbf{r}_{2}\right)  \cdots\left(  d\mathbf{r}_{d+1}\right)
\,\mathbf{r}_{23}^{i_{2}}\mathbf{r}_{23}^{j_{2}}\cdots\mathbf{r}_{dd+1}%
^{i_{d}}\mathbf{r}_{dd+1}^{j_{d}}\,\left|  \psi\left(  \mathbf{r}%
_{d+1}\right)  \cdots\psi\left(  \mathbf{r}_{2}\right)  \right|  ^{2}%
\]%
\[
B_{i}=-2k\delta_{j_{1}j_{2}\cdots j_{d}}^{ii_{2}\cdots i_{d}}\,%
{\textstyle\idotsint}
\left(  d\mathbf{r}_{2}\right)  \cdots\left(  d\mathbf{r}_{d+1}\right)
\,\mathbf{r}_{2}^{j_{1}}\,\mathbf{r}_{23}^{i_{2}}\mathbf{r}_{23}^{j_{2}}%
\cdots\mathbf{r}_{dd+1}^{i_{d}}\mathbf{r}_{dd+1}^{j_{d}}\,\left|  \psi\left(
\mathbf{r}_{d+1}\right)  \cdots\psi\left(  \mathbf{r}_{2}\right)  \right|
^{2}%
\]%
\[
Z=k\delta_{j_{1}j_{2}\cdots j_{d}}^{i_{1}i_{2}\cdots i_{d}}\,%
{\textstyle\idotsint}
\left(  d\mathbf{r}_{2}\right)  \cdots\left(  d\mathbf{r}_{d+1}\right)
\,\mathbf{r}_{2}^{i_{1}}\mathbf{r}_{2}^{j_{1}}\,\mathbf{r}_{23}^{i_{2}%
}\mathbf{r}_{23}^{j_{2}}\cdots\mathbf{r}_{dd+1}^{i_{d}}\mathbf{r}%
_{dd+1}^{j_{d}}\,\left|  \psi\left(  \mathbf{r}_{d+1}\right)  \cdots
\psi\left(  \mathbf{r}_{2}\right)  \right|  ^{2}%
\]
It is now straightforward to use the field equations
\[
i\frac{\partial}{\partial t}\psi\left(  \mathbf{r}\right)  +\frac{1}{2m}%
\nabla^{2}\psi\left(  \mathbf{r}\right)  =\left(  Z+\mathbf{r}^{i}%
B_{i}+\mathbf{r}^{i}\mathbf{r}^{j}K_{ij}\right)  \psi\left(  \mathbf{r}%
\right)
\]
to determine a closed set of time-derivative equations obeyed by the
coefficients in the effective potential. \ Once these are solved, either in
special situations or perhaps more generally, then the field equation itself
is to be solved using the time-dependence determined for the coefficients, in
a self-consistent way. \ Sounds easy, even if it is not in practice,
but\ perhaps the resulting non-isotropic equations can always be solved in
closed form in some limit, such as large $n$.

There was neither enough time in the talk nor enough space in this written
version to discuss either the supersymmetric extensions of these models or
their quantization using deformation methods \cite{CPZ,TLCunpub}. \ These
subjects will be treated elsewhere.\bigskip

\noindent{\large Acknowledgments}\newline This work was supported in part by
NSF Award 0073390 and by US Department of Energy, Division of High Energy
Physics, Contract W-31-109-ENG-38. \ I thank Cosmas Zachos and the Particle
Theory Group at Argonne National Laboratory for their hospitality in the
summer of 2001 during which a portion of this research was completed.

\end{document}